\documentclass[aps,preprint,showpacs,superscriptaddress,groupedaddress]{revtex4}  
\usepackage{graphicx}  
\usepackage{dcolumn}   
\usepackage{bm}        
\usepackage{amssymb}   
\usepackage{amsmath}
\usepackage{titlesec}
\usepackage[font=small,labelfont=bf]{caption}
\newcommand{\be}{\begin{equation}}
\newcommand{\ee}{\end{equation}}

\begin{document}
\title{Higgs Inflation and General Initial Conditions}

\author{Sarang Zeynizadeh}\email{zeynizadeh@physics.sharif.ir}\affiliation{Department of Physics, Sharif University of Technology, Tehran, Iran}\affiliation{School of physics, Institute for Research in Fundamental Sciences (IPM), Tehran, Iran}
\author{Amin Rezaei Akbarieh}\email{am_rezaei@physics.sharif.ir}\affiliation{Department of Theoretical Physics, Tabriz University, Tabriz, Iran}\affiliation{Research Institute for Astronomy and Astrophysics of Maragha (RIAAM), Maragha 55134-441, Iran}
\date{\today}

\begin{abstract}
 Higgs field of particle physics  can play the role of the inflaton in the early universe, if it is non-minimally coupled to gravity. The Higgs inflation scenario  predicts  a small  tensor to scalar  ratio: $r\simeq 0.003$. Although this value  is consistent with the upper bound $r<0.12$ given by BICEP2/\emph{Keck Array} and Planck data, but it is not at their maximum likelihood point: $r\simeq 0.05$. inflationary observables depend not only on the inflationary models, but also depend on the initial conditions of inflation. Changing initial state of inflation can improve the value of $r$. In this work, we study the Higgs inflation model under general initial conditions and show that there is a subset of these general initial conditions which leads to enhancement of $r$. Then we show that this region of parameter space is consistent with non-Gaussianity bound.     
\end{abstract} 

\maketitle
\section{I\MakeLowercase{ntroduction}}
Inflationary epoch of early universe has became an important part of standard big bang  model of cosmology \cite{Linde:2014nna}. Inflationary  paradigm not only solve two basic problems of standard cosmology, i.e. the horizon and flatness problems, but also predicts that the large-scale structure of universe originated from the primordial perturbation is nearly scale invariant which is in good agreement with the observation \cite{Ade:2015lrj,Ade:2014xna}.

 In single field models of inflation generally a scalar field which is called inflaton, drives an exponentially expansion. It will be economical if we identify a known particle with the inflaton field.  The Higgs field of particle physics has the chance to be identified with the inflaton field. The first model constructed on this assumption, has been proposed by Bezrukov and Shaposhnikov \cite{Bezrukov:2007ep}. They claimed that the Higgs field  can be identified with the inflaton field if it is non-minimally coupled to gravity. At the same time, dimensionless coupling constant $\xi$ which will be defined in section two, should be of order $10^4$. This large value of coupling constant leads to the unitarity violation \cite{Barbon:2009ya}.  Unitarity violation implies that there should be a UV cut off $\Lambda$. Beyond $\Lambda$ our effective theory will be broken down. Hence our theory should be replaced by a new fundamental theory for beyond UV cut off.   Another issue related to the model in \cite{Bezrukov:2007ep},  is very small value prediction for the tensor to scalar perturbation ratio $r$, i.e. $r\simeq 0.003$. Announcement of BICEP2 for B-mode detection \cite{Ade:2014xna} with a large value $r\simeq 0.2$ motivated people to do some efforts to reconcile Higgs inflation with BICEP2 results \cite{Hamada:2014iga}. Soon, it  turns out that there is serious doubt about BICET2 results \cite{Flauger:2014qra,Mortonson:2014bja}. Recently, \emph{Keck Array} acclaim that they also have found an excess of  B-mode power over the standard expectation which is consistent with the BICEP2 results \cite{Ade:2015fwj}.  In a joint analysis, BICEP2/\emph{Keck Array} and Planck collaborations report their results as a likelihood curve for $r$ with an upper limit $r<0.12$ and a maximum likelihood at $r\simeq 0.05$ \cite{Ade:2015tva}.   Although the predicted value for $r$ by Higgs inflation model is consistent with upper limit in \cite{Ade:2015tva}, but still is far enough from the maximum likelihood  $r\simeq 0.05$. Therefore, it is reasonable to search for some way to increase the $r$ value. In the presence of UV cut off which in turn introduce a new physics, $r$ can be altered due to the non-trivial initial state effects. Hence,  $r\simeq 0.003$ is not the firm prediction of Higgs inflation model, but it will depend on the initial state of inflation.
 
Determination of initial conditions is the necessary condition to describe dynamics of a given system. When there is no UV cut off, initial condition is trivial and choosing Bunch-Davies vacuum satisfied Minkowski space limit. In the presence of UV cut off, the effects of new theory can be sit on the non-trivial initial condition or non-Bunch-Davies vacuum \cite{Easther:2001fi}\cite{Anderson:2005hi}. Here we are going to show that by suitable choice of initial condition, it is possible to get sizeable value for $r$. Recently, Ashoorioon   
\emph{et al.} \cite{Ashoorioon:2013eia} employ the non-trivial initial conditions for the chaotic model of inflation  to suppress the value of $r$ to reconciliation of the  Planck data with that of  the BICEP2. They exclude a large piece of parameter space by using the observational bound of the non-Gassianity. But in the case of the Higgs inflation model this exclusion does not need to occur due to the special property of the Higgs inflation model. Incidentally, using this region  leads to the enhancement of $r$. 
  
The paper is organized as follows. In section II we give a brief review of Higgs inflation model proposed in \cite{Bezrukov:2007ep}. In section III with a very concise review of perturbation theory in cosmology, we just mention the effects of general initial condition on the infaltionary parameters. In section IV we argue that for some region of parameter space, we obtain sizeable value for  $r$. Finally we discuss about the results and summarize them.

\section{R\MakeLowercase{eview of \MakeUppercase{H}iggs inflation}} 
To write the  standard model of particle physics in presence of gravity, the Higgs inflation model is one of our choices. In this model the Higgs field non-minimally coupled to the gravity via a dimensionless coupling constant $\xi$ \cite{Bezrukov:2007ep}:
\be\label{action} 
L=L_{SM}-\frac{1}{2}M^2R-\xi H^\dagger HR,
\ee 
where $M$ is some mass scale, $R$ and $H$ denote the Ricci scalar and Higgs field respectively. 
The potential term is required by renormalizibility of the scalar field in a curved background. 
By choosing unitary gauge $H=h/\sqrt{2}$, scalar sector non-minimally coupled to gravity:  
\be\label{action jordan}
S_J=\int d^4x\sqrt{-g}\left[-\frac{1}{2}M^2R-\frac{1}{2}\xi h^2 R+\frac{1}{2}\partial_\mu h\partial^\mu h-V(h)\right],
\ee 
where sub-index $J$ indicates Jordan frame,  and the  potential, $V(h)$, is defined: 
\be\label{potential1}
V(h)=\frac{1}{4}\lambda(h^2-v^2)^2,
\ee
where $v=\left\langle h\right\rangle $. For $1\ll\xi\ll10^{17}$, we can assume $M\simeq M_p$ where $M_p$ is the reduced Planck mass. Due to the presence of the  non-minimal coupling term, it is very cumbersome to work with. By following conformal transformation, we can transform the Jordan frame to the Einstein frame:   
\be\label{confrmal }
g_{\mu\nu}\to \hat{g}_{\mu\nu}=\Omega^2g_{\mu\nu},\qquad
\qquad \Omega^2=1+\frac{\xi h^2}{M^2_p}
\ee   
This transformation leads to a non-canonical kinetic term which  can be converted  to canonical form by field redefinition
\be\label{field redefinition1}
\frac{d\chi}{dh}=\sqrt{\frac{\Omega^2+6\xi^2 h^2/M^2_p}{\Omega^4}}
\ee
The action in Einstein frame becomes
\be\label{action einstein}
S_E=\int d^4x \sqrt{-g}\left\lbrace -\frac{1}{2}M^2_p\hat{R}+\frac{1}{2}\partial_\mu\chi\partial^\mu\chi-U(\chi)\right\rbrace 
\ee
where $\hat{R}$ is Ricci scalar in terms of $\hat{g}_{\mu\nu}$. The potential term  $U(\chi)$ is
\be\label{potential2}
U(\chi)=\frac{1}{\Omega^4}\frac{\lambda}{4}\left(h(\chi)^2-v^2\right)^2
\ee
According to (\ref{confrmal }) and (\ref{field redefinition1}), for large values of $h$, i.e. $h\gg M_p/\sqrt{\xi}$, we have:
\be\label{field redefinition2}
h\simeq\frac{M_p}{\sqrt{\xi}}\exp\left(\frac{\chi}{\sqrt{6}M_p}\right)
\ee
\be\label{potential3}
U(\chi)=\frac{\lambda M^4_p}{4\xi^2}\left(1+\exp\left(-\frac{2\chi}{\sqrt{6}M_p}\right)\right)^{-2}
\ee
 for large values of $h$ or $\chi\gg\sqrt{6}M_p$, the potential $U(\chi)$ is flat. Wherein the  Higgs field drives inflation. 
In order to show that whether this potential can gives an consistent inflationary expansion, we use the standard slow roll formalism in Einstein frame
\be\label{slow roll}
\epsilon =\frac{1}{2}M^2_p\left(\frac{U'}{U}\right)^2,\quad \eta =M^2_p\frac{U''}{U},\quad N=\int \frac{1}{\sqrt{2\epsilon}}\frac{d\chi}{M^2_p},\quad  n_s=1-6\epsilon +2\eta,\quad r=16\epsilon
\ee
where $\epsilon$ and $\eta$ are slow roll parameters, $N$ is the number of e-folding, $n_s$ denotes the spectral index and $r$ is the tensor to scalar perturbation ratio. Substituting  (\ref{potential3}) in (\ref{slow roll})  and considering large field values for $h$ leads to \cite{Bezrukov:2013fka}
\be\label{epsilon}
\epsilon\simeq \frac{4M^4_p}{3\xi^2 h^4},
\ee
\be\label{eta}
\eta\simeq -\frac{4M^2_p}{3\xi h^2},
\ee
 \be\label{e-folding}
 N\simeq \frac{6}{8}\frac{\xi}{M^2_p}\left(h^2_N-h^2_{\text{end}}\right),
 \ee
where $h_N$ denotes the field value at the horizon exit and $h_{\text{end}}$ denotes the field value at the end of inflation. End of inflation corresponds to $\epsilon=1$. Using equation (\ref{epsilon}) we obtain $h_{\text{end}}\simeq \frac{1.07M_p}{\sqrt{\xi}}$.
The $N$ is determined from the CMB observation: $N\simeq57.7$ \cite{Bezrukov:2013fka}. Substituting this value in (\ref{e-folding}) leads to $h_N\simeq \frac{9.14M_p}{\sqrt{\xi}}$. 
From the observation \cite{Ade:2013zuv} we can put constraint on the amplitude of the scalar power spectrum 
\be\label{scalar power spectrum}
\Delta_{\cal R}^2=\frac{1}{8\pi^2}\frac{H^2}{\epsilon M^2_p}\simeq 2\times 10^{-9}.
\ee
This constraint can be used to determine unknown parameter $\xi$. By using (\ref{epsilon}) and (\ref{scalar power spectrum}) in the slow roll regime, $U\simeq 3M^2_p H^2$,  we obtain
\be\label{u/epsilon}
\frac{U}{\epsilon}=24\pi^2 M^4_p\Delta^2_{\cal R}\simeq (0.027M_p)^4.
\ee  
Evaluating $\frac{U}{\epsilon}$ at $h_N$, equation (\ref{u/epsilon}) gives rise to \cite{Bezrukov:2013fka}
\be\label{xi}
\xi=47000\sqrt{\lambda}.
\ee
Using (\ref{e-folding}), $n_s$ and $r$ evaluated at $h_N$ can be approximated  as 
\be\label{ns and r}
n_s\simeq 1-8\frac{4N+12}{(4N+3)^2},\qquad r\simeq \frac{192}{(4N+3)^2}.
\ee
Where $N\simeq 57.7$ gives $n_s\simeq 0.967$ and $r\simeq 0.0031$ \cite{Bezrukov:2013fka}.

\section{P\MakeLowercase{rimordial perturbation  with general initial condition}}
In this section we will review the cosmological perturbation theory to realise how  the effects of general initial conditions come into the game. We will  just mention their effects on inflationary quantities such as power spectrum, spectral index, etc. In order to derive the equations governing the perturbation,  we consider a minimally coupled scalar field with arbitrary potential \cite{Baumann:2009ds}
\be\label{action} 
S=\int d^4x\sqrt{-g}\left[\frac{1}{2}M^2_pR-\frac{1}{2}g^{\mu\nu}\partial_\mu\phi\partial_\nu\phi-V(\phi)\right].
\ee
Perturbations are defined around the homogeneous background given by the solutions of the action (\ref{action}), i.e. $\bar{\phi}(t)$ and $\bar{g}_{\mu\nu}(t)$
\be\label{perturbation}
\phi(t,\bold{x})=\bar{\phi}(t)+\delta\phi(t,\bold{x}),\qquad g_{\mu\nu}(t,\bold{x})=\bar{g}_{\mu\nu}(t)+\delta g_{\mu\nu}(t,\bold{x})
\ee
and the perturbed metric is parametrised as 
\be\label{perturbed metric}
ds^2=a^2(\tau)\left[-(1+2\Phi)d\tau^2+((1-2\Psi)\delta_{ij}+h_{ij})dy^idy^j\right].
\ee
Where $\tau$ is the conformal time, $a$ is the scalar factor of FRW metric, $\Phi$ and $\Psi$ are Bardeen potentials and $h_{ij}$  denotes a symmetric tensor with $h^i_i=0,\, \partial^i h_{ij}=0$. In addition to physical degrees of freedom, these perturbations also can contain the fictitious gauge freedom. To avoid these gauge freedom, it is useful to introduce a new gauge invariant scalar quantity 
\be\label{comoving curvature perturbation}
{\cal R}=\Psi +\frac{H}{\dot{\bar\phi}}\delta\phi.
 \ee
Which is called comoving curvature perturbation. Expanding action in (\ref{action}) up to second order in terms of ${\cal R}$ leads to
\be\label{second order action}
S_{(2)}= \frac{1}{2}\int d^4x a^3 \frac{\dot{\phi}^2}{H^2}\left[\dot{{\cal R}}^2-a^{-2}(\partial_i{\cal R})^2\right].
\ee
By defining the Mukanov-Sasaki variable
\be\label{mukhanov-sasaki variable}
v\equiv z{\cal R}, \qquad z^2\equiv a^2\frac{\dot{\phi}^2}{H^2}, 
\ee
equation of motion corresponding with second order action, becomes
\be\label{mukhanov equation}
v''_k+\left(k^2-\frac{z''}{z}\right)=0.
\ee
Where $v_k$ is Fourier mode of $v$ and prime indicates derivative with respect to conformal time. 
For quasi-de sitter background in slow-roll limit, general solution of the equation (\ref{mukhanov equation}) can be written as  
\be\label{general solution}
v_k\simeq\frac{\sqrt{\pi \lvert\tau\rvert}}{2}\left[\alpha^S_k H^{(1)}_{\frac{3}{2}}(k\lvert\tau\rvert)+\beta^S_k H^{(2)}_{\frac{3}{2}}(k\lvert\tau\rvert)\right].
\ee
Where $H^{(1)}_{\frac{3}{2}}$ and $H^{(2)}_{\frac{3}{2}}$ are  the first and second kind Hankel functions respectively. $\alpha^S_k$ and $\beta^S_k$ are Bogoliubov coefficients that satisfy Wronskian constraint 
\be\label{wronskian constraint}
\lvert \alpha^S_k\rvert^2-\lvert\beta^S_k\rvert^2=1
\ee
Since $\alpha^S_k$ and $\beta^S_k$ are arbitrary up to Wronskian constraint, they correspond to the general initial condition.  In the case of  $\alpha^S_k=1$ and $\beta^S_k=0$, $v_k$ corresponds to the standard  BD vacuum. The states with generic values of  $\alpha^S_k$ and $\beta^S_k$ usually are called non-BD  vacuum or $\alpha$-vacua. 
Dimensionless scalar  power spectrum is defined 
\be\label{scalar power spectrum definition}
\Delta^2_S=\frac{k^3}{2\pi^2}\left\lvert\frac{v_k}{z}\right\rvert^2_{k=aH}
\ee
Substituting  (\ref{general solution}) into (\ref{scalar power spectrum definition}) leads to \cite{Ashoorioon:2013eia,Boyanovsky:2006qi}
\be\label{gamma S}
\Delta^2_S=\frac{1}{8\pi^2\epsilon}\left(\frac{H}{M_p}\right)^2 \gamma_S,\qquad \gamma_s=\left\lvert\alpha^S_k-\beta^S_k\right\rvert^2_{k=aH}
\ee
Non-Gaussianity as an important probe of the early universe encodes in bispectrum. Having power spectrum in squeezed $k_3\ll k_1\sim k_2$ limit suffices to obtain bispectrum.  According to \cite{Kundu:2013gha}, three point function of scalar perturbation in squeezed limit  for $\alpha$-vacua  is given by
\be\label{bispectrum}
\left\langle {\cal R}_{\bold k_1}{\cal R}_{\bold k_2}{\cal R}_{\bold k_3}\right\rangle \simeq(2\pi)^3\delta(\bold k_1+\bold k_2+\bold k_3)\left[4\epsilon\left(\frac{k_1}{k_3}\right)\Phi(k_1, k_3)-6\epsilon+2\eta\right]P_{\cal R}(k_1)P_{\cal R}(k_3).
\ee  
Where $P_{\cal R}(k)=\frac{2\pi^2}{k^3}\Delta^2_S$ and
\be\label{Phi}
\Phi(k_1, k_3)= 2\text{Re}\left[\alpha^S_{k_1}\beta^S_{k_1}\left(\frac{\alpha^{S*}_{k_1}-\beta^{S*}_{k_1}}{\alpha^{S}_{k_1}-\beta^{S}_{k_1}}\right)\left(\frac{\alpha^{S}_{k_3}+\beta^{S}_{k_3}}{\alpha^{S}_{k_3}-\beta^{S}_{k_3}}\right)\right].
\ee
Noting  equation (\ref{bispectrum}), local non-Gaussianity parameter $f_\text{NL}^\text{local}$ becomes \footnote{for definition of $f_{\text{NL}}^{\text{local}}$ refer to \cite{Baumann:2009ds}.}
\be\label{fnl}
f_\text{NL}^\text{local}\simeq\frac{5}{12}\left[4\epsilon\left(\frac{k_1}{k_3}\right)\Phi(k_1, k_3)-6\epsilon+2\eta\right]
\ee
In this paper, since we are only interested in the local configuration of  non-Gaussianity, we don't need to consider flattened and equilateral configurations.
Similarly for tensor perturbations, we obtain the following mode function  
\be\label{tensor mode function}
h_k(\tau)\simeq\frac{\sqrt{\pi\lvert\tau\rvert}}{2}\left[\alpha^T_k H^{(1)}_{\frac{3}{2}}(k\lvert\tau\rvert)+\beta^T_k H^{(2)}_{\frac{3}{2}}(k\lvert\tau\rvert)\right].
\ee
Dimensionless tensor power spectrum also has a similar relation \cite{Ashoorioon:2013eia,Boyanovsky:2006qi}
\be\label{tensor power spectrum}
 \Delta^2_T=\frac{2}{\pi^2}\left(\frac{H}{M_P}\right)^2\gamma_T,\qquad \gamma_T=\left\vert\alpha^T_k-\beta^T_k\right\vert^2_{k=aH}.
 \ee
Where $\alpha^T_k$ and $\beta^T_k$ are  Bogoliubov coefficients  and satisfy Wronskian condition $\rvert\alpha^T_k\rvert^2-\lvert\beta^T_k\rvert^2=1$. The tensor to scalar perturbation ratio $r$ in the case of $\alpha$-vacua is given by  \cite{Ashoorioon:2013eia}
\be\label{r for alpha}
r=\frac{\Delta^2_T}{\Delta^2_S}=16\epsilon \gamma,\qquad \gamma=\frac{\gamma_T}{\gamma_S}
\ee
\section{e\lowercase{nhancement of $r$ for  some $\alpha$-vacua}}
 According to (\ref{r for alpha}) for a given $\epsilon\ll 1$, to have sizeable value of $r$, it is required that $\gamma\gg 1$. If  we would like to have a large value of $r$, we have to search for some special regions in the parameter space of initial conditions, such that their corresponding  $\gamma$, satisfies $\gamma\gg 1$. In the case of Higgs inflation model with a small $r$, we can use some special initial conditions with $\gamma\gg 1$ to rise the $r$ to the large value. Due to the arbitrariness of $\alpha_k$ and $\beta_k$ up to Wronskian condition, we may think that it is possible to obtain a $\gamma$ as large as we would like. In fact, there is an upper limit on the value of $\gamma$ due to  the constraints from the back reaction effects and observational  bound on the non-Gaussianity. 

 Back reaction effects should be small enough  not to destroy the inflationary background.  By assuming the crude model $\beta_k\sim\beta_0 e^{-\frac{k^2}{M^2 a^2}}$, this condition gives rise to \cite{Holman:2007na}
\be\label{back reaction}
\lvert\beta_0\rvert\leq\sqrt{\epsilon\lvert\eta\rvert}\,\frac{M_p H}{M^2}.
\ee
Where we assumed $\epsilon\ll\eta$,  that is reasonable assumption in the case of Higgs inflation. $M$ in (\ref{back reaction})  is an energy scale of the new physics  and our effective theory is valid only in energies lower than the energy scale $M$. In order to convert  the constraint on $\beta_0$ in (\ref{back reaction}) to a constraint on $\gamma$, we are following the notation of \cite{Ashoorioon:2013eia}. It is mentioned in \cite{Ashoorioon:2013eia} that $\gamma_S$ and $\gamma_T$ depend on the  relative phases of $\alpha_k$ and $\beta_k$. Therefore  it is useful to  parametrise them as 
\be\label{alpha and beta scalar }   
\alpha_k^S=\cosh\chi_S e^{i\varphi_S},\qquad \beta^S_k=\sinh \chi_S e^{-i\varphi_S},
\ee
\be\label{alpha and beta tensor}
\alpha_k^T=\cosh\chi_T e^{i\varphi_T},\qquad \beta^T_k=\sinh \chi_T e^{-i\varphi_T}.
\ee
Consistency of above parametrization with  $\beta_k\sim\beta_0 e^{-\frac{k^2}{M^2 a^2}}$ implies that
\be\label{consistency of parametrization} 
\lvert\beta_0\rvert e^{-\frac{k^2}{M^2 a^2}}=\sinh \chi.
\ee  
Below the energy scale of new physics, $k<a M$, $e^{-\frac{k^2}{a^2M^2}}\simeq 1$ and (\ref{consistency of parametrization}) becomes
\be\label{consistency of parametrization 2} 
\lvert\beta_0\rvert\simeq\sinh \chi.
\ee   
In Higgs inflation model, (\ref{epsilon}) and (\ref{eta}) results in $\epsilon\simeq 1.8\times 10^{-4}$ and $\eta\simeq -1.6\times10^{-2}$. Thus, using $\Delta^2_S\simeq 2\times10^{-9}$ in (\ref{gamma S}) gives $\frac{H}{M_p}=5.3\times 10^{-6}/\sqrt{\gamma_S}$. Substituting  this result into (\ref{back reaction}) and using (\ref{consistency of parametrization 2}), leads to
\be\label{scalar back reaction}
\frac{M^2}{H^2}\lesssim 323 \frac{\sqrt{\gamma_S}}{\sinh \chi_S}.
\ee
There is a similar expression for tensor modes except that $\chi_S$ is replaced by $\chi_T$. Physical expectations implies $M>H$. Let us  write $\gamma$ in terms of new parametrization in (\ref{alpha and beta scalar }) and (\ref{alpha and beta tensor}) 
\be\label{gamma}
\gamma=\left\vert \frac{\cosh \chi e^{i\varphi_T}-\sinh \chi e^{-i\varphi_T}}{\cosh \chi e^{i\varphi_S}-\sinh \chi e^{-i\varphi_S}}\right\vert ^2.
\ee
Where we take $\chi_S=\chi_T=\chi$ for convenient. According to this formula, $\gamma$ can be bigger or smaller than one, e.g. for  $\chi\simeq 1$, $\varphi_S\simeq 0.01$ and $\varphi_T\simeq\frac{\pi}{2}$,  we get $\gamma\simeq 70$. In \cite{Ashoorioon:2013eia} it has been shown that for $\chi\gtrsim1$, $\varphi_S\simeq\frac{\pi}{2}$ and for generic values $\varphi_T$, we obtain $\gamma\lesssim 1$. Authors of \cite{Ashoorioon:2013eia} put away the $\varphi_S$'s that satisfy $\varphi_S\lesssim \frac{\pi}{10}$ because these values violate the observation bound on the non-Gaussianity. However  as we will show, in the Higgs inflation model because of smallness of slow-roll parameter $\epsilon$, we do not encounter with such a violation.

\begin{figure}   
\includegraphics[scale=1]{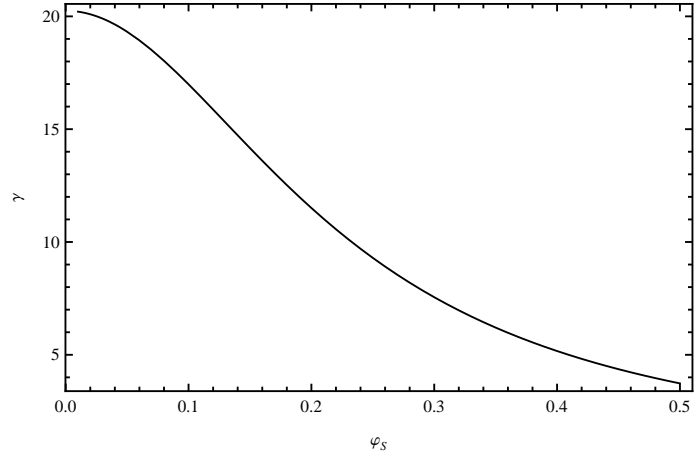}
\caption{Diagram of $\gamma$ for $\chi\simeq 0.75$ and $\varphi_T\simeq\frac{\pi}{2}$.}\label{fig-r}
\end{figure}

Basically,  enhancement of non-Gaussianity due to general initial condition can occur  only for two type of non-Gaussianity: \emph{flattened configuration} and \emph{local configuration}. 
In flattened configuration, $k_1+k_2 \simeq k_3$, enhancement finally disappear due to the effect of projection on the CMB surface \cite{Holman:2007na}. In local configuration, $k_3\ll k_1+k_2$, there is no cancellation   and   enhancement of  non-Gaussianity will be survived. The effects of general initial conditions on the parameter of local non-Gaussianity i.e. $f_{NL}^{\text{local}}$,  are given by $\Phi(k_1,k_3)$  in (\ref{Phi}).  To constrain $\Phi$, let us first write (\ref{Phi}) in terms of the parameters introduced in (\ref{alpha and beta scalar })
\begin{eqnarray}\label{Phi2}
\Phi(k_1, k_3)=2 \text{Re}\left[\cosh\chi\sinh \chi\left(\frac{\cosh \chi e^{-i\varphi_S}-\sinh\chi e^{i\varphi^S}}{\cosh\chi e^{i\varphi_S}-\sinh \chi e^{-i\varphi_S}}\right)\left(\frac{\cosh\chi e^{i\varphi_S}+\sinh\chi e^{-i\varphi^S}}{\cosh\chi e^{i\varphi_S}-\sinh\chi e^{-i\varphi_S}}\right)\right].
\end{eqnarray} 
Where $\sinh\chi(k_3)\simeq\sinh \chi(k_1)$. Because at horizon crossing, $k=aH$, (\ref{consistency of parametrization 2}) implies $\sinh \chi(k)\sim e^{-\frac{H^2}{M^2}}$, thus $k$ dependence of $\sinh\chi(k)$ is lost  for $H<M$. Using Planck data \cite{Ade:2015ava}
\be\label{fnl constraint}
-4.2\lesssim f_{NL}^{\text{local}}\lesssim 5.8,
\ee
combing (\ref{fnl}) and (\ref{fnl constraint}) leads to 
\be\label{Phi constraint}
-14\lesssim\Phi\lesssim 19.
\ee
\begin{figure}   
\includegraphics[scale=1]{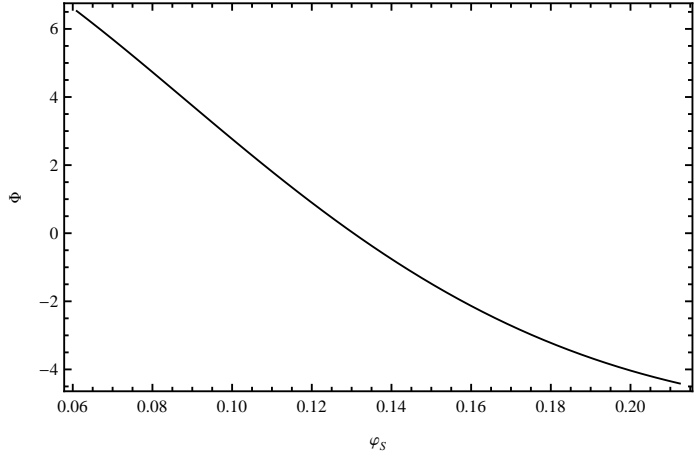}
\caption{Diagram of $\Phi$ for $\chi\simeq 0.75$.}\label{fig-phi}
\end{figure}
Where we take $ \frac{k_1}{k_3}\simeq 10^2$ \cite{Flauger:2013hra}. 
Smallness of $\epsilon$ in Higgs inflation model in comparison to it's value in other inflationary model (such as chaotic inflation model in which $\epsilon\simeq 0.01$) allows us to have large value for $\Phi$ in (\ref{fnl}), whiles $f_{NL}^{\text{local}}$ is still in the region specified in (\ref{fnl constraint}). The possibility of having large value of $\Phi$ is equivalent to the possibility of having very small value for $\varphi_S$ (Fig.\ref{fig-phi}). Small values of $\varphi_S$ provide the chance of reaching the large values of  $\gamma$. For instance, taking $\varphi_T\simeq \frac{\pi}{2}$, $\varphi_S\simeq 0.1$ and $\chi\simeq 0.75$ results in $\gamma\simeq 17$, $\Phi\simeq 3$ (Fig.\ref{fig-r}). These values by noting (\ref{scalar back reaction}) lead to $M\simeq 13H$ where is consistent with $M>H$. Using $\gamma\simeq 17$ in (\ref{r for alpha}) gives improved value $r\simeq 0.05$ . 
It should be noticed that although $\alpha$-vacua as the initial state of the system for some region of parameter space increases the tensor to scalar perturbation ratio, however it does not affect the spectral index. The spectral index is defined
\be\label{spectral index}
n_s-1=\frac{d \,\text{ln} \Delta^2_S}{d\,\text{ln}k}=\frac{d}{d\,\text{ln}k}\left[\text{ln}\frac{1}{8\pi^2\epsilon}\left(\frac{H}{M_p}\right)^2\right]+\frac{d\,\text{ln}\gamma_S}{d\, \text{ln}k},    
\ee
using (\ref{gamma S}) and (\ref{alpha and beta scalar }), second term in (\ref{spectral index}) for $\chi\simeq 0.75$, can be written as
\be\label{second term}
\frac{d\,\text{ln}\gamma_S}{d\, \text{ln}k}\simeq 2 \cot \varphi_S\frac{d \varphi_S}{d \,\text{ln}k}.
\ee
Since $\varphi_S$ is arbitrary parameter, we can assume that $\varphi_S$ be $k$ independent. With the choice, the second term in  (\ref{spectral index}) which represents the effects of $\alpha$-vacua, will be vanished and in consequence spectral index remains intact.
  
\section{C\MakeLowercase{oncluding remarks}}
 In this work we studied the Higgs inflation model under general initial conditions. The general initial conditions affect inflationary observables such as power spectrum, non-Gassianity, etc. The effects of general initial condition are constrained by the requirement that they should not spoil the inflationary background. Moreover observational bound on the non-Gaussianity of primordial perturbations gives another constraint on these initial condition effects. We argued that for some region of parameter space in initial condition, it is possible to enhance the tensor to scalar perturbation ratio, $r$. This enhancement was possible, because the Higgs inflation scenario gives a very small value for the slow-roll parameter $\epsilon$. The smallness of $\epsilon$ makes us able to access to more extent region of parameter space without violation of observation bound on the non-Gaussianity. Suitable choice of the region of parameter space can be led to a value of $r\simeq 0.05$, which is desirable value according to the latest results from the BICEP2/\emph{Keck Array} and Planck collaborations.            
\begin{acknowledgements}
S. Z would like to thank M. M. Sheikh-Jabbari and  M. Golshani for encouragement and support. This work has been supported financially by the Research Institute for Astronomy and Astrophysics of Maragha (RIAAM) under research project "Higgs inflation", Maragha, Iran.

\end{acknowledgements}


\begin{thebibliography}{99}
\bibitem{Linde:2014nna} 
  A.~Linde,
  arXiv:1402.0526 [hep-th].
\bibitem{Ade:2015lrj} 
  P.~A.~R.~Ade {\it et al.}  [Planck Collaboration],
  arXiv:1502.02114 [astro-ph.CO].
\bibitem{Ade:2014xna} 
  P.~A.~R.~Ade {\it et al.}  [BICEP2 Collaboration],
  Phys.\ Rev.\ Lett.\  {\bf 112}, 241101 (2014)
  [arXiv:1403.3985 [astro-ph.CO]].
\bibitem{Flauger:2014qra} 
  R.~Flauger, J.~C.~Hill and D.~N.~Spergel,
  JCAP {\bf 1408}, 039 (2014)
  [arXiv:1405.7351 [astro-ph.CO]].
\bibitem{Mortonson:2014bja} 
  M.~J.~Mortonson and U.~Seljak,
  JCAP {\bf 1410}, no. 10, 035 (2014)
  [arXiv:1405.5857 [astro-ph.CO]].
 \bibitem{Ade:2015fwj} 
  P.~A.~R.~Ade {\it et al.}  [BICEP2 and Keck Array Collaborations],
  arXiv:1502.00643 [astro-ph.CO].
  \bibitem{Bezrukov:2007ep} 
  F.~L.~Bezrukov and M.~Shaposhnikov,
  Phys.\ Lett.\ B {\bf 659}, 703 (2008)
  [arXiv:0710.3755 [hep-th]].
\bibitem{Ade:2015tva} 
  P.~A.~R.~Ade {\it et al.}  [BICEP2 and Planck Collaborations],
  Phys.\ Rev.\ Lett.\  {\bf 114}, no. 10, 101301 (2015)
  [arXiv:1502.00612 [astro-ph.CO]].
\bibitem{Barbon:2009ya} 
  J.~L.~F.~Barbon and J.~R.~Espinosa,
  Phys.\ Rev.\ D {\bf 79}, 081302 (2009)
  [arXiv:0903.0355 [hep-ph]].
  C.~P.~Burgess, H.~M.~Lee and M.~Trott,
  JHEP {\bf 1007}, 007 (2010)
  [arXiv:1002.2730 [hep-ph]].
   M.~P.~Hertzberg,
  JHEP {\bf 1011}, 023 (2010)
  [arXiv:1002.2995 [hep-ph]].
  C.~Germani and A.~Kehagias,
  Phys.\ Rev.\ Lett.\  {\bf 105}, 011302 (2010)
  [arXiv:1003.2635 [hep-ph]].
   F.~Bezrukov, A.~Magnin, M.~Shaposhnikov and S.~Sibiryakov,
  JHEP {\bf 1101}, 016 (2011)
  [arXiv:1008.5157 [hep-ph]].
   S.~Ferrara, R.~Kallosh, A.~Linde, A.~Marrani and A.~Van Proeyen,
  Phys.\ Rev.\ D {\bf 83}, 025008 (2011)
  [arXiv:1008.2942 [hep-th]].
\bibitem{Hamada:2014iga} 
  Y.~Hamada, H.~Kawai, K.~y.~Oda and S.~C.~Park,
  Phys.\ Rev.\ Lett.\  {\bf 112}, 241301 (2014)
  [arXiv:1403.5043 [hep-ph]].
  P.~Ko and W.~I.~Park,
  arXiv:1405.1635 [hep-ph].
  J.~Rubio and M.~Shaposhnikov,
  Phys.\ Rev.\ D {\bf 90}, 027307 (2014)
  [arXiv:1406.5182 [hep-ph]].
   Y.~Hamada, H.~Kawai, K.~y.~Oda and S.~C.~Park,
  arXiv:1408.4864 [hep-ph].
  \bibitem{Easther:2001fi} 
  R.~Easther, B.~R.~Greene, W.~H.~Kinney and G.~Shiu,
  Phys.\ Rev.\ D {\bf 64}, 103502 (2001)
  [hep-th/0104102].
  \bibitem{Anderson:2005hi} 
  P.~R.~Anderson, C.~Molina-Paris and E.~Mottola,
  Phys.\ Rev.\ D {\bf 72}, 043515 (2005)
  [hep-th/0504134].
\bibitem{Ashoorioon:2013eia} 
  A.~Ashoorioon, K.~Dimopoulos, M.~M.~Sheikh-Jabbari and G.~Shiu,
  JCAP {\bf 1402}, 025 (2014)
  [arXiv:1306.4914 [hep-th]].
 
A.~Ashoorioon, K.~Dimopoulos, M.~M.~Sheikh-Jabbari and G.~Shiu,
  Phys.\ Lett.\ B {\bf 737}, 98 (2014)
  [arXiv:1403.6099 [hep-th]].
\bibitem{Bezrukov:2013fka} 
  F.~Bezrukov,
  Class.\ Quant.\ Grav.\  {\bf 30}, 214001 (2013)
  [arXiv:1307.0708 [hep-ph]].
\bibitem{Ade:2013zuv} 
  P.~A.~R.~Ade {\it et al.}  [Planck Collaboration],
  Astron.\ Astrophys.\  {\bf 571}, A16 (2014)
  [arXiv:1303.5076 [astro-ph.CO]].




  
\bibitem{Baumann:2009ds} 
  D.~Baumann,
  arXiv:0907.5424 [hep-th].
\bibitem{Boyanovsky:2006qi} 
  D.~Boyanovsky, H.~J.~de Vega and N.~G.~Sanchez,
  Phys.\ Rev.\ D {\bf 74}, 123006 (2006)
  [astro-ph/0607508].
\bibitem{Kundu:2013gha} 
  S.~Kundu,
  JCAP {\bf 1404}, 016 (2014)
  [arXiv:1311.1575 [astro-ph.CO]].
  \bibitem{Flauger:2013hra} 
  R.~Flauger, D.~Green and R.~A.~Porto,
  JCAP {\bf 1308}, 032 (2013)
  [arXiv:1303.1430 [hep-th]].
  
  
  
  
  
  
  
   
\bibitem{Holman:2007na} 
  R.~Holman and A.~J.~Tolley,
  JCAP {\bf 0805}, 001 (2008)
  [arXiv:0710.1302 [hep-th]].

\bibitem{Ade:2015ava} 
  P.~A.~R.~Ade {\it et al.}  [Planck Collaboration],
  arXiv:1502.01592 [astro-ph.CO].


\end{thebibliography}
\end{document}